\documentclass[useAMS,usenatbib]{mn2e}
\usepackage{txfonts}
\usepackage{graphicx}
\usepackage{textcomp}
\usepackage{natbib}
\usepackage{deluxetable}

\newcommand{\um}{\textmu m }
\newcommand{\uu}{\textmu m}

\newcommand{\tausil}{$\rm{\tau_{9.7}}$ }
\newcommand{\tausilu}{$\rm{\tau_{9.7}}$}
\newcommand{\ssil}{$S\!_{\rm{sil}}$ }
\newcommand{\ssilu}{$S\!_{\rm{sil}}$}

\newcommand{\lsixtwo}{L$_{\rm{62}}$}

\newcommand{\lir}{$L_{\rm{IR}}$}

\newcommand{\csil}{$C_{\rm{7-10-14}}$ }
\newcommand{\csilu}{$C_{\rm{7-10-14}}$}
\newcommand{\ssildecomp}{$S^{*}\!\!\!_{\rm{sil}}$ }
\newcommand{\fsil}{$F_{\rm{sil}}$}
\newcommand{\fcont}{$F_{\rm{cont}}$}

\bibpunct[]{(}{)}{;}{a}{}{,}

\title[An atlas of mid-infrared spectra of galaxies]{An atlas of mid-infrared spectra of star-forming and active galaxies}

\author[Hern\'an-Caballero \& Hatziminaoglou]{
A. Hern\'an-Caballero$^{1,2}$\thanks{e-mail: ahc@denebola.org}
and E. Hatziminaoglou$^3$\\
$^{1}$Instituto de Estructura de la Materia, Consejos Superior de Investigaciones Científicas, Serrano 121, 38006 Madrid, Spain\\
$^{2}$Centro de Astrobiolog\'ia (CSIC/INTA), Ctra. de Torrej\'on a Ajalvir km 4, 28850 Torrej\'on de Ardoz, Spain\\
$^{3}$European Southern Observatory, Karl-Schwarzschild-Str. 2, 85748 Garching bei M\"unchen, Germany\\
}

\begin{document}

\date{Accepted ........ Received ........;}

\maketitle

\label{firstpage}

\begin{abstract}

We present a panoramic atlas of {\it Spitzer}/IRS spectra of extragalactic sources collected from the
recent literature, with value added measurements of their spectral features obtained in a homogeneous 
and concise manner. The
atlas covers the full spectrum of the extragalactic universe and includes star forming galaxies, 
obscured and unobscured active galaxies, luminous and ultra-luminous infrared galaxies, and hybrid
objects. Measured features such as the polycyclic aromatic hydrocarbons, the strength of the silicates
in emission or absorption around 9.7 \uu, rest-frame monochromatic luminosities or colours, combined
with measurements derived from spectral decomposition, are used
to establish diagnostics that allow for classification of sources, based on their infrared
properties alone. Average templates of the various classes are also derived.
The full atlas with the value added measurements and ancillary archival data are publicly available
at http://www.denebola.org/atlas, with full references to the original data.

\end{abstract}

\section{Introduction}

The mid-to-far infrared spectral range holds the key to the physical processes that shape the life
and properties of
galaxies, namely star formation and nuclear activity. This spectral domain, and with the
exception of the very bright emission lines of the brightest sources, was almost unaccessible
from the ground and until 15 years ago, when the sensitivity and large wavelength coverage of
the Infrared Space Observatory (ISO) allowed the detailed study of large numbers of nearby starburst
galaxies and active galactic nuclei (AGN) and lead to the first mid-infrared (MIR) diagnostic tools.
ISO studies observed, among other things, the presence of polycyclic aromatic hydrocarbon (PAH) features 
in the majority of star-forming galaxies \citep[]{Lutz98, Rigopoulou99} 
and their more scarce appearance in AGN \citep[e.g.][]{Sturm00}, 
as well as the presence of a silicate feature in absorption in type 2 AGN \citep{Clavel00}.

Though the studies carried out with ISO brought a break through in our understanding of star formation
in galaxies, it was the {\it Spitzer} Space Telescope with its MIR cameras IRAC and MIPS, and the 
InfraRed Spectrograph (IRS) that revolutionised
our knowledge on the starburst and AGN phenomena and their connection.
IRAC broad band MIR photometry alone allowed for a rough classification among AGN and a distinction
between active and star-forming galaxies
\citep[e.g.][]{Lacy04, Stern05, Hatziminaoglou05, Hatziminaoglou09}, while IRS observations of
hundreds of extragalactic sources provided with more elaborate diagnostic tools \citep[e.g.][]{Spoon07} and 
showed that their MIR spectra are richer and more complex than once assumed.
The new discoveries, however, raised new issues that models are now challenged to reproduce.
The silicate emission feature in type 1 AGN peaks at longer wavelengths than the 9.7 \um predicted by the models 
\citep[e.g.][]{Siebenmorgen05, Hao05}; silicate emission has also been observed in 
type 2 AGN \citep{Sturm06, Teplitz06, Nikutta09} in apparent contradiction to at least some of the suggested dust
geometries; and PAH emission is
present in many quasars \citep[e.g.][]{Schweitzer06}, despite previous theories postulating their destruction in the
vicinity of the active nucleus. Meanwhile, accurate measurements of PAH features in star-forming galaxies
are made difficult by the presence of other features, such as the
deep silicate absorption at 9.7 \um or an underlying AGN continuum.

Understanding the processes linking different galaxy populations and finding their distinctive characteristics
requires the comparison of the properties of large samples, selected with different but well understood criteria.
Studies of individual samples usually pursue specific goals, and
basic MIR properties for diagnostics (such as the strength of the silicate features and PAH bands)  
are measured in ways adapted to the needs of each individual study, but not necessarily consistently across samples, 
making a direct comparison of results painful, if not impossible.

With this in mind, we compiled a list of 739 {\it Spitzer}/IRS spectra of extragalactic sources,
the largest of its kind to date, previously scattered in the recent literature. The sample, though not statistical,
comprises a variety of objects, namely type 1, 2 and intermediate AGN, ULIRGs, starburst and sub-mm galaxies, 
with a large span of spectral properties such as PAH and silicate features, continua and distinct physical characteristics.
The MIR properties were meassured in a concise way for all objects and can thus be directly compared.
The paper is structured as follows.
\S2 describes the sample selection, overall properties and collection of ancillary data; \S3
presents the extraction of the IRS spectra and calibration procedure followed; sections 4 and 5 detail the methods used to 
measure MIR spectral properties and the composite spectra obtained for the various subsamples, respectively; \S6
reproduces a few common diagnostic diagrams and finally \S7 summarizes our conclusions.

The full collection of IRS spectra with references to the original data, ancillary data and additional 
measurements are available at http://www.denebola.org/atlas.

\section{Sample selection}

The sample hereby presented is drawn from papers in the literature presenting 
{\it Spitzer}/IRS spectroscopy of active and star-forming galaxies, and contains a wide variety of targets including 
Starbursts, Seyferts, LINERs, ULIRGs, QSOs, SMGs and radio galaxies, among others. A total 739 unique sources
are selected from 
\citet{Brand08}, 
\citet{Brandl06}, 
\citet{Buchanan06},
\citet{Cao08}, 
\citet{Dasyra09}, 
\citet{Deo07}, 
\citet{Farrah09}, 
\citet{Haas05}, 
\citet{Hao05}, 
\citet{Hernan-Caballero09}, 
\citet{Hiner09}, 
\citet{Imanishi07, Imanishi10},
\citet{Imanishi09}, 
\citet{Lacy07}, 
\citet{Leipski09},
\citet{Lutz08}, 
\citet{Maiolino07}, 
\citet{Menendez-Delmestre09}, 
\citet{Murphy09}, 
\citet{Netzer07}, 
\citet{Polletta08}, 
\citet{Pope08}, 
\citet{Shi06},
\citet{Siebenmorgen05}, 
\citet{Sturm05, Sturm06}, 
\citet{Tommasin08}, 
\citet{Valiante07}, 
\citet{Weedman05, Weedman06a, Weedman06b, Weedman09},
\citet{Willett10},
\citet{Wu09},
\citet{Yan07} and
\citet{Zakamska08}.

Table \ref{IRS_observations_data} summarizes the type and number of spectra and observation mode for each
subsample. For a more detailed description of the selection criteria and properties of the individual subsamples
we defer the reader to the corresponding papers.

{\it Spitzer}/IRS data for sources in the sample include observations with both the Low-Resolution (R$\sim$100) and
High-Resolution (R$\sim$600) IRS modules, all performed in staring-mode, except for the sources from \citet{Buchanan06}
and \citet{Wu09}, where observations were performed in spectral mapping mode, but only the spectrum obtained with the 
slit placed in the galactic nucleus is given.
Maximum spectral coverage is 5-38\um in the Low-Res module and 10-35\um in the High-Res one, but in many faint targets
only a subset of Low-Res modules is applied, resulting in a reduced spectral coverage that is not uniform even 
within the same subsample.

Sources at intermediate and high redshift appear unresolved to the IRS, and thus the spectra integrate the
emission of the whole galaxy. By contrast, in nearby targets the spectra only sample the nuclear region.
Our selection does not put constraints on redshift, not even the existence of a consistent redshift estimate,
but individual subsamples usually target certain redshift range, or indirectly bias the selection through flux or
colour constraints.

Spectroscopic redshifts (either from optical or MIR spectroscopy) are available for 
695 of the 739 sources. Redshifts are obtained from the literature and cross-checked with the NASA Extragalactic 
Database (NED).
525 redshifts come from optical spectra, while 170 sources (mostly obscured high-$z$ targets) rely on a redshift
estimate from their IRS spectrum. In targets where both optical and MIR spectroscopic redshifts are available,
priority is given to the former, except in cases of strong discrepancy in which the optical redshift is flagged 
as unreliable or the MIR spectrum shows very clear PAH bands. 
The remaining 44 sources have no published spectroscopic redshift, or it is
uncertain. They are not removed from the sample, but will be ignored in the subsequent analysis.

\begin{figure} 
\begin{center}\hspace{-0.3cm}
\includegraphics[width=8.5cm]{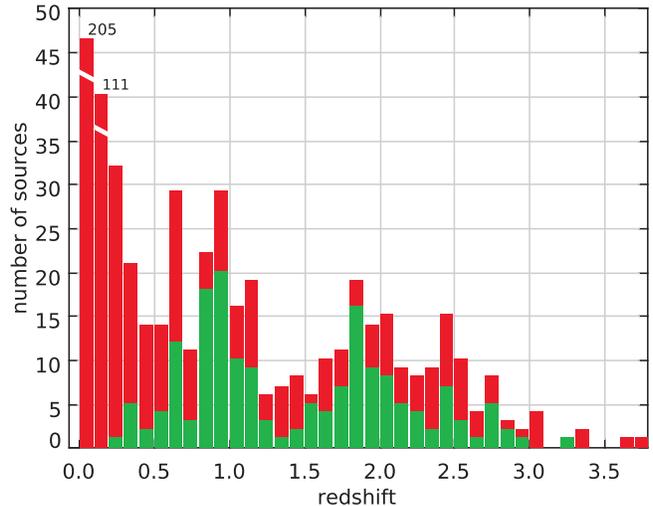}
\end{center}
\caption[Redshift histogram]{Redshift histogram for the 695 sources with spectroscopic redshifts in the
sample. The red bars indicate optical redshifts, while the green bars represent the MIR ones. Bars for the first two redshift bins are truncated to improve readability.\label{zhistogram}}
\end{figure}

\begin{figure} 
\begin{center}\hspace{-0.3cm}
\includegraphics[width=8.5cm]{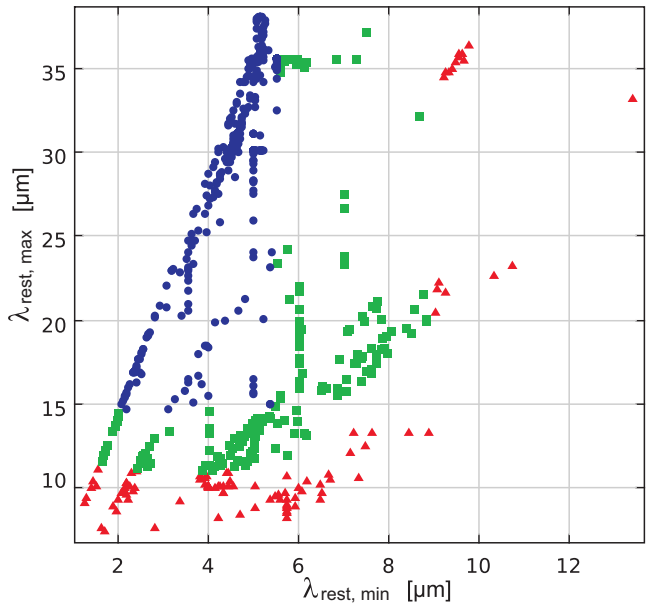}
\end{center}
\caption[Restframe wavelength coverage]{Restframe wavelength coverage for the 695 sources with spectroscopic
redshift in the sample. Each point represents one source, with the X and Y coordinates indicating the 
shortest and longest wavelengths sampled in the IRS spectrum, respectively. 
Symbols indicate spectral coverage in the 5--15\um restframe range: $>$95\% (blue circles), between 60\% and 95\% (green boxes) and $<$60\% (red triangles).\label{lambdarest_coverage}}
\end{figure}

About half of the sample, 347 sources, have redshifts below z=0.3, while the rest spans a large redshift
range up to z$\sim$3.7 (Figure \ref{zhistogram}). 
Spectral coverage in rest frame wavelength is in all cases within 1.2 and 38 \uu, with 389 sources sampling
the 5.5-14 \um range in its entirety and 550 the critical 8-12\um range, which contains both the 10 \um silicate
feature and 11.3 \um PAH band (see Figure \ref{lambdarest_coverage}).

An optical spectroscopic classification is also available for most sources. A positional match with the 13th version
of the Veron Catalogue of Quasars \& AGN \citep[VERONCAT;][]{Veron06} yields 361 matches; for these sources, we adopt the activity type
designation in the VERONCAT. For the remaining ones we adopt the optical classification in NED, when available,
or at the very least a broad class (e.g.: ULIRG, SMG) either from NED or the individual papers the spectra are taken from.
Some sources are put in more than one classes. For example, a starburst galaxy with a Seyfert 2 nucleus will retain both
starburst and Seyfert 2 clasifications.

To deal with the large variety of spectral types, we group them into 10 categories, namely: Seyfert 1 (Sy1), Seyfert 2
(Sy2), intermediate Seyfert type (Sy1.X), LINER, Quasar (QSO), type 2 Quasar (QSO2), Starburst, ULIRG, SMG and ``other"
when none of the above classifications apply.

We set a threshold value of $\nu$L$_\nu$(7\uu) = 10$^{44}$ erg s$^{-1}$
(roughly the output of the least luminous ULIRG of this sample), above which all type 1 and type 2 AGN 
are considered as QSO or QSO2, while keeping the Seyfert 1 and Seyfert 2 designations for the lower luminosity Seyferts.
The LINER class does not include LINER cores of ultraluminous and starburst galaxies, as their MIR spectrum is typically
dominated by the host galaxy.

\section{Extraction of the spectra}

To obtain the IRS spectra we have not gone through the normal process of querying the {\it Spitzer} archive,
reduce the raw Spitzer/IRS data and extract the spectra. Instead, we have relied on Postscript figures 
contained in papers submitted to the arxiv.org preprint archive. This approach has several shortcomings 
that we describe later, and is not optimal for a detailed analysis of individual targets. Nevertheless,
it provides a cost-efficient means for performing statistical studies on the largest starburst and active
galaxy sample compiled to date.  

Sets of points representing the spectra in the figures are transformed from Postscript coordinates 
to wavelength and flux values using software developed by the authors. 
Paper figures represent the spectra in many different fashions: the independent variable is usually wavelength,
whether in the observed or rest frame, but it can also be frequency. For the dependent variable there are several
values of choice: f$_\nu$, f$_\lambda$, $\nu$f$_\nu$, $\nu$L$_\nu$, etc., that in addition are expressed in
different units systems and plotted in linear or logarithmic scale.

To homogenize the sample, we convert all spectra back to the same units provided by the
{\it Spitzer} IRS reduction pipeline: observed wavelength in micron and spectral flux density in Jy.
The accuracy with which 
the original wavelength and flux values are recovered is limited by the resolution of the Postscript figure;
nevertheless, the uncertainty that this introduces to the wavelength calibration is about an order of magnitude 
smaller than the spectral resolution (R$\sim$100) in the Low-Resolution module of IRS, and thus its impact is negligible. 

Since the uncertainties in the flux density of the spectra are not shown in most of the figures, we devised
a procedure to estimate them based on individual exposure times, obtained from queries to the {\it Spitzer} archive when they were not
indicated in the corresponding papers, using the SPEC-PET exposure time calculator. 
Flux uncertainties estimated by SPEC-PET are corrected for wavelength sampling (SPEC-PET assumes R=50), with an additional $\sqrt2$
factor to account for the noise introduced by standard background subtraction.
We use two different methods to evaluate the accuracy of these flux uncertainty estimates: i) compare with the known uncertainties 
in the spectra from papers where this information is available: \citet{Hernan-Caballero09}, \citet{Yan07}, \citet{Menendez-Delmestre09}
and \citet{Murphy09}; ii) compare with the RMS noise of the continuum in spectra that are plotted without smoothing
\citep[e.g.][]{Weedman05,Lutz08,Pope08,Zakamska08}.

Both approaches indicate that our flux uncertainty estimates are correct within a factor 2 in most sources. Even if a higher 
accuracy would be desirable, the statistical results drawn from the sample are little, if at all, affected. For an in-depth analysis of
individual targets authors should use instead the latest version of the data products
available at the {\it Spitzer} archive.

\section{Measurements}

We provide measurements for several MIR spectral features that are commonly used to estimate the
star-formation or nuclear activity, their relative contribution to the energy output of extragalactic sources 
or the amount of obscuration; namely PAH features, the 10 \um silicate feature and rest-frame monochromatic luminosities.  

\subsection{PAH features}

PAH emission is a key tracer of star formation,
with the luminosity of the PAH bands frequently used as an indirect estimator of the 
Star Formation Rate \citep[SFR; e.g.][]{Rigopoulou99, Peeters04, Brandl06, Farrah08} 
while their ratios relate to the ionization field in the interstellar medium
\citep{Draine01, Rapacioli05, Brandl06, Galliano08}.
For nearby galaxies with strong PAH bands and high signal-to-noise (S/N) 
spectra, it is possible to perform a detailed modelling of the
continuum and Lorentzian profiles of the PAH bands \citep[e.g.][]{Smith07,Treyer10}. 
For higher redshift sources, a similar analysis can be performed, provided adaptations are made to account for AGN 
emission and dust extinction \citep[e.g.][]{Sajina07,Hernan-Caballero09}, but as the equivalent width (EW) of the feature,
the S/N of the spectrum and the wavelength coverage decrease, degeneracy among solutions quickly becomes a major source of 
uncertainty.
Simpler methods such as spline or linear interpolation of the local continuum and integration 
in the PAH band \citep[e.g.][]{Genzel98,Rigopoulou99,Brandl06,Spoon07} are much more resistant to weak or noisy features, but significantly underestimate
PAH luminosities due to losses at the wings of the Lorentzian profiles and overlap between adjacent PAH bands \citep{Smith07,Treyer10}.

As already mentioned, the basic idea behind this work is to apply the same procedure when measuring the
various features across all samples. 
We therefore opted to use an intermediate approach, in which the continuum is locally interpolated and the feature flux 
is integrated in a narrow band, but both measurements are later corrected for the
expected width and shape of the feature's profile. 

The measurement proceeds as follows: we select two narrow, continuum bands at both sides of each PAH feature,
and estimate the average flux, f$_\nu$, in these bands. We perform a linear interpolation to estimate the continuum underlying the
feature, and subtract it from the spectrum. Finally we integrate the residual in a band centered at the expected wavelength of
the peak of the PAH feature to obtain the integrated PAH flux.
In order to maximize the S/N in the integrated PAH flux and to reduce the uncertainty in the underlying continuum, we
use a rather narrow integration band (which loses a significant fraction of the PAH flux in the wings of the profile) 
and place the continuum bands in its close vicinity, without trying to avoid contamination by the wings of the PAH feature
(see parameter values in Table \ref{tabla_pahs}).

The bias introduced by this procedure is compensated by a correction factor on the assumption that the PAH profile is Lorentzian with a known
full width half maximum (FWHM). 
We estimate the error that this procedure introduces by meassuring line and continuum fluxes in a set of model continuum+Lorentzian spectra
covering a wide range of EW, FWHM, and S/N values. We find that a 10\% increase in the FWHM causes a 5-6\% drop in the PAH flux,
with no observable dependency on the EW or S/N of the spectrum. 

Uncertainties in the PAH flux and the underlying continuum for each source are estimated by performing Monte Carlo simulations. 
Finally, we estimate EWs of the PAH features dividing the integrated PAH flux by the interpolated 
continuum at the center of the feature, and convert PAH fluxes to luminosities assuming a concordance $\Lambda$CDM 
cosmology ($\Lambda=0.7$, $\Omega=0.3$ and $H_0=70$ km s$^{-1}$ Mpc$^{-1}$).

Since this approach is not valid for the overlapping 7.7 and 8.6 PAH bands, we perform measurements only for the 6.2 and 11.3 \um 
PAH features (which incidentally are the most used for diagnostics). The 6.2 \um feature is measured in 587 sources, 
that at 11.3 \um in 577, and both of them together in 481 sources. 
 
\subsection{Silicates\label{silicates}}

The 10 \um silicate feature, which can appear in emission or absorption, provides insight into the geometry of 
the dust distribution along the line of sight of the star-forming regions or the active nucleus as well as
the amount of obscuration \citep[e.g.][]{Imanishi03,Sturm05,Imanishi07}.
As with the PAH bands, the main difficulty in estimating the strength of the 10 \um silicate 
feature is the identification of the underlying continuum. In sources with weak or no PAH
emission, we interpolate linearly in log($\lambda$), log(f$_\lambda$) between anchor points
at restframe 8.1 and 14.0 \uu; but whenever there is significant PAH emission ($r_{\rm{PDR}}$ $>$ 0.1, see \S\ref{decomposition}), 
we substitute the first 
anchor point by 5.55 \um in order to avoid the distortion introduced by the wings of the
7.7 and 8.6 \um PAH features.

We subtract from the spectrum the interpolated continuum, and integrate the residual (silicate
feature) in the 9-11 \um range. 
The sign of this quantity allows to determine whether the feature is in
emission (integral $>$ 0) or in absorption (integral $<$ 0). 
The theoretical wavelength of the silicate peak is 9.7 \uu, but it is known that in many 
quasars and Seyfert galaxies the feature peak is shifted to longer wavelengths when it 
appears in emission \citep[e.g.][]{Siebenmorgen05,Netzer07}. In addition, many star-forming sources show strong
emission in the S(3) 9.665 \um transition of molecular hydrogen, which contaminates the
peak of the silicate absorption profile.

Since many spectra are too noisy to accurately measure where the peak of the silicate feature actually
occurs, we simply assume it is located at $\lambda_p$ = 9.8 \um for sources with silicate absorption 
and $\lambda_p$ = 10.5 \um for silicate emission spectra. 
A detailed analysis of the 10 and 18 \um silicate features in the AGN-dominated sources
of this sample, including measurement of the peak's wavelength, is the topic of a specialised study
(Hernan-Caballero et al. in preparation).

The feature's peak flux density is used to calculate the silicate
strength, \ssil = ln[F($\lambda_p$)/F$_C$($\lambda_p$)]. As in the PAH bands, 
uncertainties in the silicate parameters are estimated using Monte-Carlo simulations.
\ssil measurements are obtained for 537 sources.

We also define the ``three-band'' colour index, \csilu, as: 

\begin{displaymath}
C_{7-10-14} = -2.5~ log \Big{(} \frac{\sqrt{f_7 f_{14}}}{f_{10}} \Big{)}
\end{displaymath}

\begin{figure} 
\begin{center}
\includegraphics[width=8.5cm]{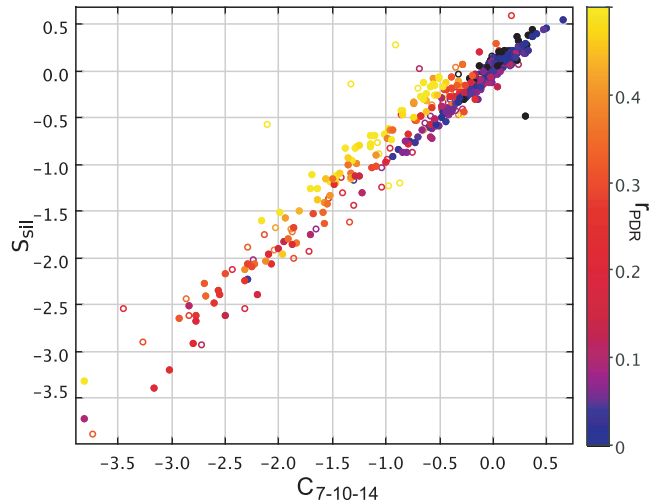}
\end{center}
\caption[\ssil vs \csilu]{Comparison between two estimates of the silicate strength: \ssilu, obtained
by the standard procedure (see text) and \csilu, calculated as a three-band colour index.
Sources with low uncertainties in the estimation of \ssil (e\ssil $<$ 0.2) are shown in filled circles,
while open circles represent the remaining objects.\label{Ssil_i7-10-14}}
\end{figure}

This parameter is an alternative, easily obtained estimator of the silicate strength, 
that shows a good correlation with \ssil in sources of all types (Figure \ref{Ssil_i7-10-14}).

In sources with little or no PAH emission, \ssil and \csil are mostly interchangeable, 
while sources with strong PAH emission show a small but consistent bias towards \csil $<$ \ssilu.
This may be produced by a steeper slope of the starburst continuum component shortwards of $\sim$8.5 \um 
that decreases the interpolated 9.8 \um continuum for \ssil (and increases thus the \ssil value),
and also from contamination by the wings of the 6.2 and 7.7 \um PAH features
in the 7.0 \um flux, that reduces the \csilu \, estimate.
The difference, $\Delta_{\rm{sil}}$ = \ssil-\csil increases with the strength of the
starburst component as measured by EW$_{6.2}$, and 
can be as high as $\sim$ 0.5 in extreme cases.

\subsection{Monochromatic luminosities}

Rest-frame monochromatic luminosities, $\nu$L$\nu$, allow for the comparison of spectral energy distributions of sources
at different redshift without the need for K-corrections. We estimate them
based on the IRS spectra at rest-frame 
wavelengths: 2.2, 3.6, 5.6, 7.0, 9.0, 10.0, 12.0, 14.0, 15.0, 18.0, 20.0 and 25.0 \uu. To this end,
each rest frame spectrum is integrated in a narrow band (between 0.5 and 2.0 \um wide depending on the wavelength) 
centered at the nominal wavelength, and divided by the band width. Table \ref{tabla_nuLnu} summarizes these measurements.

Rest frame MIR colours and colour indices are also calculated and tabulated.
Such quantities give a quick but accurate idea of the shape of a spectrum.
The [5.6-15] \um colour, for instance, is a proxy for MIR continuum slope, while the three-band
7-10-14 \um colour index (\csilu) can be used as a substitute estimate of the silicate strength
in noisy or poorly sampled spectra (see \S\ref{silicates}).

\section{Spectral Decomposition}
\label{decomposition}

Decomposition of MIR spectra
in several spectral components is a powerful tool that can provide considerable insight into the physics of the 
sources \citep[e.g.][]{Laurent00, Tran01}. 

MIR emission from active and star-forming galaxies arises mostly from HII regions, Photodissociation Regions (PDRs) and
hot dust heated by energetic X-ray-to-UV photons from an AGN, so we have used a simple model comprising the
superposition of three spectral templates (HII, PDR and AGN) obscured by a screen of dust.

The spectral decomposition is performed by fitting the 5--15 \um restframe range 
of the spectrum to a parametrized $F_{\lambda}(\lambda)$ of the form: 

\begin{equation}
F_{\lambda}(\lambda) = e^{-b \tau(\lambda)} \Big{(}a_{1} f_{\rm{AGN}}(\lambda) + 
a_{2} f_{\rm{HII}}(\lambda) + a_{3} f_{\rm{PDR}}(\lambda) \Big{)}
\end{equation}

\noindent where the free parameters ($b$, $a_1$, $a_2$, $a_3$) are calculated using a 
Levenberg--Marquardt $\chi^2$-minimization algorithm. A single optical depth value is applied to the
three spectral components, but for most sources the solution is almost identical when two
optical depth values (one for the AGN and other for the PDR and HII components) are used,
with an increased $\chi^2$ value due to the extra free parameter.

$\tau$($\lambda$) is obtained from the Galactic Centre extinction law \citep{Chiar06}, while
$f_{\rm{AGN}}$ is represented by the IRS spectrum of the PG QSO 3C273 \citep{Wu09},
$f_{\rm{PDR}}$ by the ISOCAM spectrum of a PDR in the reflection nebula NGC 7023 \citep{Cesarsky96a} and
$f_{\rm{HII}}$ by the ISOCAM spectrum of M17 in the vicinity of OB stars \citep{Cesarsky96b}.
Uncertainties in the fit parameters are estimated using Monte-Carlo simulations.

\begin{figure} 
\begin{center}
\includegraphics[width=8.5cm]{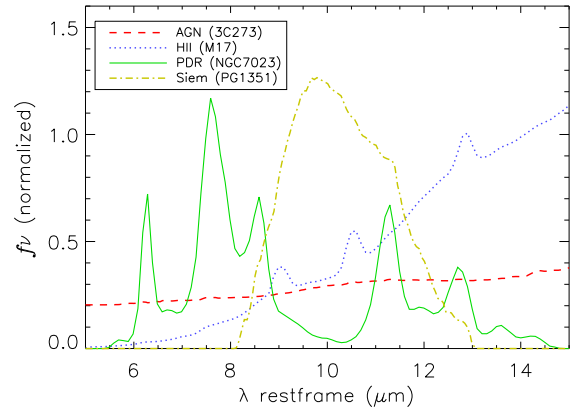}
\end{center}
\caption[templates decomposition]{5--15\um spectra of the AGN, HII and PDR templates used in the spectral decomposition.
The silicate emission template used in the alternative spectral decomposition model for silicate emission sources is also shown.\label{templates_decomp}}
\end{figure}

\begin{figure}
\begin{center}
\includegraphics[width=9cm]{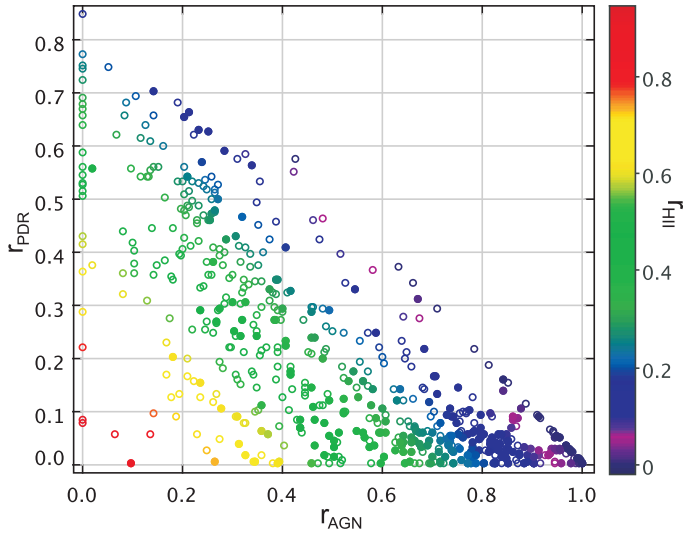}
\end{center}
\caption[rAGN_rPDR]{Contribution of PDR and AGN components to the integrated 5--15\um restframe luminosity as derived from the spectral decomposition. Filled symbols indicate sources with full 5.5-15 \um coverage in the IRS spectrum, while open symbols represent sources with only partial coverage. The amount of HII component can be calculated by the formula: $r_{\rm{HII}}$ = 1 - ($r_{\rm{AGN}}$+$r_{\rm{PDR}}$).\label{rAGN_rPDR}}
\end{figure}

We quantify the contribution, $r$, of each spectral component to the MIR spectrum as the ratio of its
integrated luminosity to the total luminosity in the 5--15 \um restframe range, $r_{\rm{AGN}}$,
$r_{\rm{HII}}$ and $r_{\rm{PDR}}$ for the AGN, HII and PDR components.
Figure \ref{rAGN_rPDR} shows the distribution of the three spectral components in the sample. 

Only 235 of the 695 sources with available redshifts have IRS spectra covering the 5--15 \um spectral range in its entirety.
In many local galaxies, the IRS spectrum starts at 5.2--5.5 \uu, while in higher redshift sources the long wavelength
end of the spectrum is often shortwards of 15 \uu. Nevertheless, 379 spectra cover more than 95\% of the 5--15 \um range, and 595 show coverage ratios above 60\%.

To validate the results on sources with incomplete spectral coverage, we have performed test runs of the decomposition
algorithm on the subsample with full 5.5--15 \um coverage, but restricting the fitting range to 5--13.5 \uu, 5--12 \uu,
and 5--10.5 \uu. For each spectral component, $i$, we define the change in the estimated share of the 5--15 \um luminosity
as: $\Delta r_i$ = $r_i'$ - $r_i$, where $r_i'$ is the value of $r_i$ obtained when fitting in the restricted range.

The results (Table \ref{test_decomp}) show that 
decreasing the long wavelength limit to 13.5 \um has almost no effect, with typical $\Delta r_i$ of a few percent,
and no significant bias. A further reduction of the long wavelength limit down to 12 \um significantly
reduces the contribution of the HII template, by 7 percentage points on average, and increases the dispersion in all components
by a factor $\sim$ 3. Nevertheless, moving the limit to 10.5 \um does not increase the bias, which improves
sligthly for the HII and PDR templates; but the dispersion further increases by another factor $\sim$ 2, to almost 0.2 for
HII, which renders the decomposition results rather unreliable.
Note that the HII component is by far the most affected by truncation of the spectrum at long wavelengths because
of its steep slope.

\begin{figure} %
\begin{center}\hspace{-0.7cm}
\includegraphics[width=9cm]{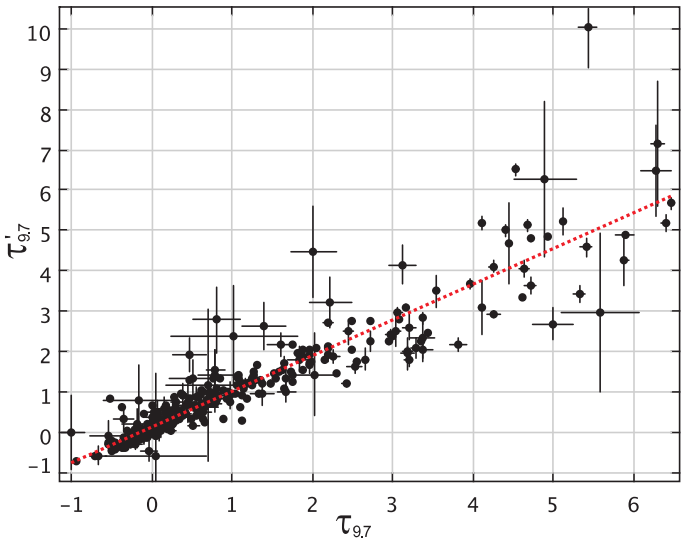}
\end{center}
\caption[tau_prima_tau]{Comparison of decomposition \tausil values obtained by trimming spectral coverage
to 5--12\um restframe (Y axis) versus \tausil obtained using
the full 5--15 \um range (X axis). The dotted line represents a linear fit to the data.\label{tau_prima_tau}}
\end{figure}

$\Delta$\tausil also shows high dispersion for a 12 \um limit, but most of the strong deviations originate in spectra
with poor S/N which show large parameter uncertainties even with full spectral coverage (Figure \ref{tau_prima_tau}).
There is no significant bias for $\Delta$\tausil except for the test in the 5--10.5 \um spectral range.

In order to obtain reliable measurements without sacrificing too many sources, we opt to require a lower limit of 60\% coverage
in the 5--10.5 \um spectral range to perform spectral decomposition. In addition, we provide coverage ratios for all sources in the sample.\\

\begin{figure} 
\begin{center}
\includegraphics[width=8.5cm]{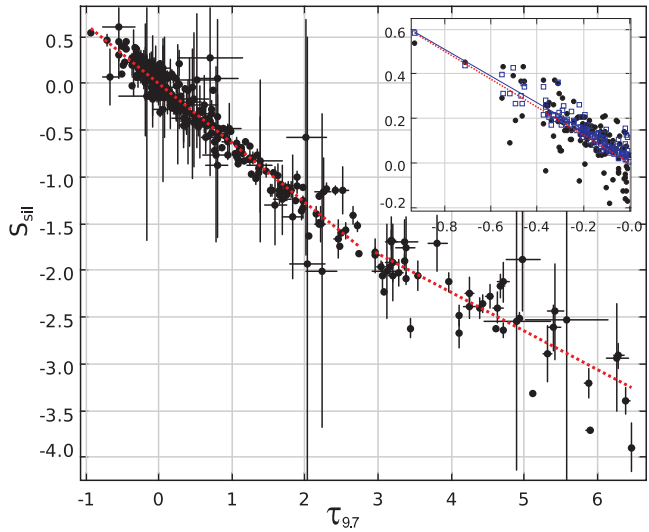}
\end{center}
\caption[Ssil_tausil]{\ssil versus \tausil for the sources in the sample with complete 5.5--15 \um spectral coverage.
Linear fits for sources with \tausil $<$ 3 and \tausil $>$ 3 are shown in dotted lines. The inset plot is a close up
to the silicate emission region of the diagram where only sources with uncertainties in \ssil below 0.2 are shown, for clarity.
The open blue squares are the \ssildecomp estimates obtained from the alternative spectral decomposition model (see text), plotted
as a function of \tausilu.\label{Ssil_tau}}
\end{figure}

Figure \ref{Ssil_tau} shows \ssil versus the optical depth at 9.7 \um estimated
from the $b$ parameter (\tausilu) for the 379 sources with $>$95\% coverage in the
5--15\um range.

The tight correlation and the significantly larger uncertainties in
\ssil compared to \tausilu \, are noteworthy. There appears to be a slope change at \ssil $\sim$ -2, (\tausil $\sim$ 3) 
that might
be caused by saturation of the silicate feature: since \ssil depends on the 9.7 \um flux,
it is difficult to obtain an accurate measurement when the flux level nears zero; also, emission
in the 9.66 \um S(3) line of molecular hydrogen (found in many ULIRGs) can increase the measured
9.7 \um flux. On the contrary, \tausil takes advantage of the whole 5--15 \um spectrum, and we
can obtain a sensible estimate of the 9.7 \um optical depth from the wings of the silicate profile,
even if the feature is saturated.\\

When the silicate feature appears in emission, the best fit is obtained for a negative
value of the opacity parameter $b$, since this produces a 10 \um bump with a shape alike that of
the silicate emission feature. This unphysical solution allows for a decent fit of silicate
emission sources with no extra parameters in the model, and the associated negative \tausil
values correlate with \ssil values (Figure \ref{Ssil_tau}).

In order to check the validity of this approach, and to investigate the influence of a negative
$b$ value in the relative contribution of the three spectral components, we compare the results
with an alternative spectral decomposition model for sources with the silicate feature in emission.
In the alternative model, we fix the opacity parameter
at $b$ = 0 and include a fourth spectral template with the silicate emission profile
of PG1351 (the source with the strongest silicate emission in the sample), obtained
as the excess flux over the interpolated continuum between 8 and 13.5 \uu.
Figure \ref{templates_decomp} shows the 5--15 \um spectrum of the four components.
The opacity needs to be fixed when using a silicate emission template to avoid degeneracies,
since an increase in the silicate emission component can be compensated by a
simultaneous increase in the opacity parameter.

To obtain a meassurement of the silicate strength
we use the combined 9.7 \um flux of the AGN, HII and PDR templates as the continuum estimate (\fcont), and the
9.7 \um flux of the fourth (silicate) template as the peak flux estimate of the silicate feature (\fsil).
The silicate strength is then calculated as \ssildecomp = ln(1 + \fsil/\fcont).

Comparison of \ssildecomp with \tausil from the original decomposition model shows a good correlation,
with significantly less dispersion than the correlation between \ssil and \tausil (Figure \ref{Ssil_tau}, inset plot).
Comparison of \ssildecomp with \ssil also yields a good correlation, with \ssildecomp $\sim$ 0.91 \ssilu.
Changes in the relative contribution of spectral components, when adding the Silicate template,
are small; in particular $r_{\rm{PDR}}$, a parameter we will
use for diagnostics in \S\ref{MIRclassification}, is
almost unaltered with a $\Delta r_{\rm{PDR}}$ averaging 0.0016 and a standard deviation of 0.012.

The alternative spectral decomposition model assumes the silicate feature peaks at $\sim$9.8 \uu,
while there are instances in our sample where the peak of the emission is shifted towards longer wavelengths. 
These \ssil estimates should, therefore, be considered as an approximation to the real values. 
A more accurate meassurement would require the ability to
shape and/or shift the silicate template in order to reproduce the diversity observed within the ATLAS sample.
This and other issues related to the silicate feature in AGN will be addressed in a forthcoming work.

\begin{figure} 
\begin{center}
\includegraphics[width=8.5cm]{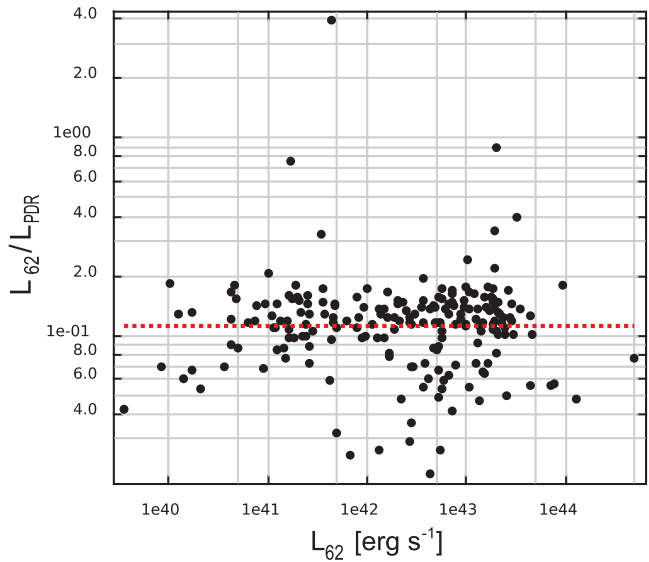}
\end{center}
\caption[LumPDR_L62]{Ratio between the 6.2 \um PAH luminosity, as obtained by integration in the 5.95--6.55 \um band with corrections for flux lost in the wings of the profile (L$_{\rm{62}}$), and the integrated PAH luminosity as derived from the PDR component in spectral decomposition (L$_{\rm{PDR}}$) as a function of L$_{\rm{62}}$ for sources with a S/N $>$ 2 detection of the 6.2 \um PAH feature. The dotted line represents the best fit linear relation.\label{LumPDR_L62}}
\end{figure}

The luminosity of the PDR component estimated in the fit (L$_{\rm{PDR}}$) is a measurement of the total PAH luminosity in the galaxy.
Comparison of L$_{\rm{PDR}}$ with the 6.2 \um PAH luminosity (Figure \ref{LumPDR_L62}) indicates the latter is responsible for roughly 10\%
of the total PAH luminosity in the galaxy, in agreement with the results of \citet{Smith07} in a sample of local starbursts, and that this ratio does not significantly change in the higher luminosity sources.

\begin{figure} 
\begin{center}
\includegraphics[width=8.5cm]{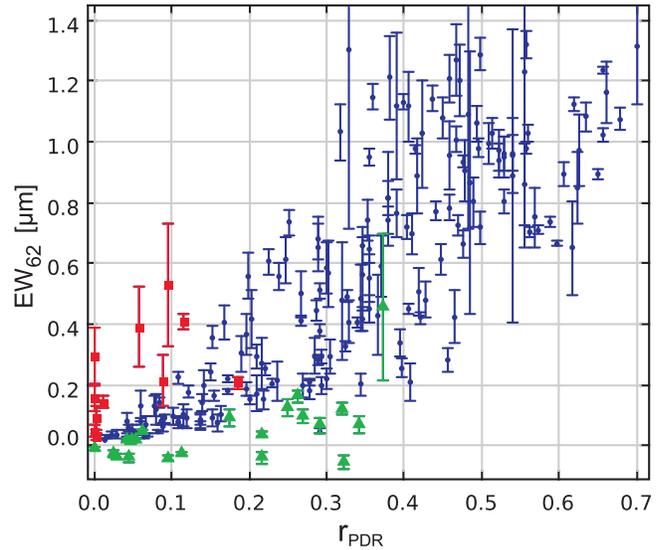}
\end{center}
\caption[EW62_rPDR]{EW$_{6.2}$ versus $r_{\rm{PDR}}$ for the sources in the sample with a S/N $>$ 2 detection of the 6.2 \um PAH feature. Symbols indicate the relative contribution of the 6.2 \um to the integrated PAH luminosity as meassured by the L$_{\rm{62}}$/L$_{\rm{PDR}}$ ratio: red squares indicate a ratio larger than 0.2, while green triangles mark ratios smaller than 0.05.\label{EW62_rPDR}}
\end{figure}

A comparison of EW$_{6.2}$ with $r_{\rm{PDR}}$ shows significant dispersion (Figure \ref{EW62_rPDR}), mostly because of variations in the relative strength of PAH features. Unusually low 6.2 \um to total PAH luminosity
ratios (L$_{6.2}$/L$_{\rm{PDR}}$ $<$ 0.05) are found mainly in heavyly obscured ULIRGs, where an absorption
band of water ice overlaps with the 6.2\um PAH emission, while very high ratios (L$_{6.2}$/L$_{\rm{PDR}}$ $>$ 0.2) appear in sources with peculiar MIR spectra that suggest unusual physical properties or artifacts in the data.

\section{MIR Diagnostics}

\subsection{The fork diagram}

The ``fork diagram'', i.e. 
the strength of the 10 \um silicate feature, \ssil, versus the equivalent width of the 6.2 \um PAH feature,
EW$_{6.2}$, first discussed by \citet{Spoon07}, is now 
considered a standard tool to distinguish between AGN- and starburst-dominated sources
\citep[e.g.][]{Farrah07,Zakamska08,Wu09,Willett10}.
Starbursts tend to have large values of EW$_{6.2}$, which decrease with increasing \ssil due to depletion
of the PAH band by absorption of adjacent water ice bands \citep{Spoon02} and increased continuum emission,
while in AGN-dominated
galaxies silicates appear in shallower absorption (typically type 2 objects) or emission (type 1 objects)
while the PAH features are less pronounced. 

\begin{figure} 
\begin{center}\hspace{-0.7cm}
\includegraphics[width=9cm]{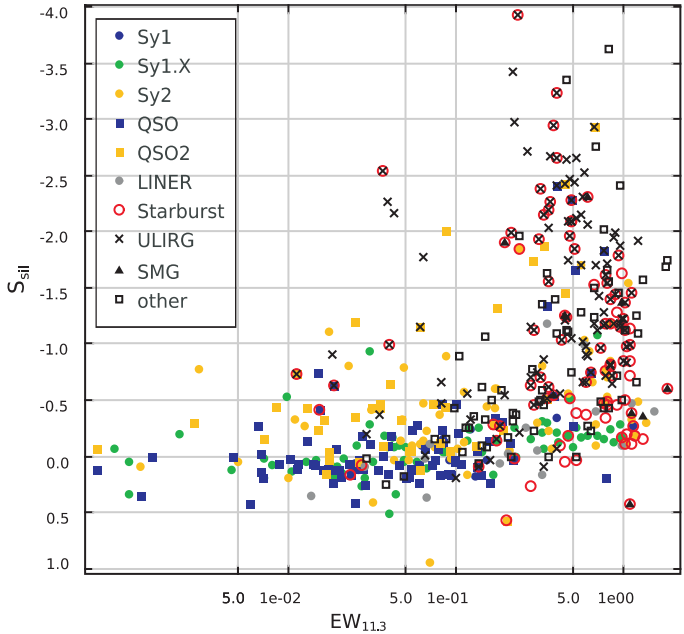}
\end{center}
\caption[Fork Diagram]{Fork Diagram showing the 10 \um silicate strength versus 
the 11.3 \um PAH equivalent width. Symbols indicate the classification of sources in the
literature. Solid symbols indicate optical AGN classification as
Seyfert 1 (blue circles), Seyfert 2 (yellow circles), intermediate Seyfert types (green
circles), LINER (grey circles), quasar (blue boxes) and type 2 quasar (yellow boxes);
red open circles indicate Starburst or HII-region type in the optical, 
while black crosses and triangles represent ULIRGs and submillimiter galaxies, respectively.
Sources not classified, or with types that do not match any of the former ones
are shown in black open squares.
\label{Ssil10_EW112}}
\end{figure}

A variant of the fork diagram can be obtained using the 11.3 \um PAH feature instead of
the 6.2 \um one (Figure \ref{Ssil10_EW112}). This approach has the advantage of requiring
less wavelength coverage and avoids the bias introduced in deeply obscured sources by water 
ice absorption in the wings of the 6.2 \um PAH feature (note that the 11.3 \um flux can be
heavyly depleted by silicate absorption, but since the silicate feature is very broad, 
the equivalent width is not significantly altered). In this variant of the diagram the starburst-dominated
branch turns nearly vertical, but the topology remains unchanged.

\begin{figure} 
\begin{center}\hspace{-0.7cm}
\includegraphics[width=9cm]{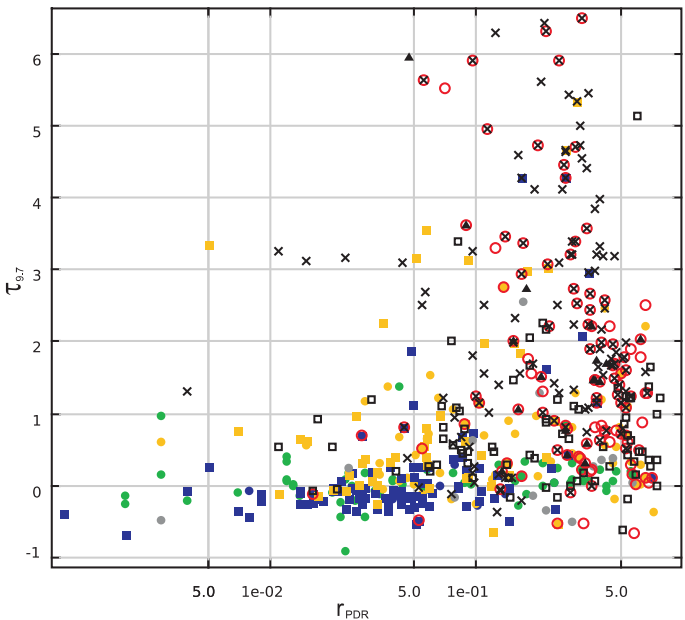}
\end{center}
\caption[Spectral Decomposition Fork Diagram]{Analog to the fork diagram obtained by representing the fraction of PDR component $r_{\rm{PDR}}$ versus the optical depth
at 9.7 \uu, \tausilu, as calculated with the spectral decomposition method. Symbols as in Fig. \ref{Ssil10_EW112}.\label{fork_decomp}}
\end{figure}

The parameters derived from the spectral decomposition allow us to build an alternative fork-like diagram,
where $r_{\rm{PDR}}$ substitutes the PAH EW and \tausil takes the place of \ssil
(Figure \ref{fork_decomp}). The advantage of this procedure is the results do not depend on
the measurement of a single PAH band and the flux at the core of the silicate feature and the related caveats; 
instead, it uses the whole 5--15 \um spectrum, making it less vulnerable to noise.  

\subsection{MIR classification\label{MIRclassification}}

We use the spectral decomposition results to obtain a MIR classification of the sources in the ATLAS
sample. At a first stage, we separate AGN-dominated and starburst-dominated sources based on
the fraction of PDR component, $r_{\rm{PDR}}$. 

We set the limit between AGN-dominated and starburst-dominated sources at $r_{\rm{PDR}}$ = 0.15,
which corresponds to EW$_{6.2}$ $\sim$ 0.2 \um according to the relation shown in Figure \ref{EW62_rPDR}.

Hernan-Caballero et al. (2009) showed that typical $\nu$L$_\nu$(5.5\uu)/\lir ratios for AGN
have values around 0.3 (their Figure 13), and starburst sources have mean \lsixtwo/\lir ratios of 0.01
(their Table 9). Aproximating the 6.2 \um continuum with the 5.5 \um luminosity would imply an equivalent
width of EW$_{6.2}$ $\sim$ 0.2 \um for a source with 50\% starburst and 50\% AGN contribution to the IR
luminosity.
 
Since many spectra are noisy and there is a significant number of sources near this threshold, 
we require the 1-$\sigma$ confindence interval not to contain the threshold value in order to qualify
as AGN-dominated or starburst-dominated. The remaining sources are classified as ``composites'' if
$r_{\rm{PDR}}$ has S/N $>$ 2 or ``unknown'' otherwise.

In 93 sources the spectral coverage in the 5--15\um range is too low to estimate $r_{\rm{PDR}}$, but the equivalent width of the 6.2 or 11.3 \um PAH features is meassured. For these sources we use EW$_{11.3}$ = 0.2 \um (or EW$_{6.2}$ = 0.2 \uu)
as an alternative boundary between the starburst-dominated and AGN-dominated classes.

With these criteria, 257 sources are classified as starburst-dominated in the MIR (MIR\_SB),
348 as AGN-dominated (MIR\_AGN) and 29 are too close to the threshold to decide, so they get 
classified as composites (MIR\_COMP). The spectra of another 49 sources are too noisy to tell, and in the
remaining 12 the 6.2 and 11.3 PAH features are both outside of the observed spectral range.

The 348 AGN-dominated sources are further sub-classified according to their \tausil meassurement into
silicate emission (MIR\_AGN1, 119 sources) or absorption (MIR\_AGN2, 160 sources), again requiring the 1-$\sigma$ confidence interval in \tausil not to cross \tausil = 0 for a clean classification. 
Sources in which the 1-$\sigma$ interval contains \tausil = 0,
but with uncertainties below 0.1 are classified as "flat MIR-spectrum AGN" (MIR\_AGNx, 18 sources) to indicate that there is no strong emission or absorption at 10 \uu, and the remaining 51 sources retain the generic MIR\_AGN classification.

Table \ref{optical_MIR_table} shows the distribution of sources in the sample according to their optical
spectroscopic and MIR classifications. Values indicate the number of sources
with a given optical and MIR classification. Sources classified in the literature
as both AGN and starburst in the optical are discarded for clarity,
but the counts that would be obtained were they included, are shown in brackets.

In most sources, the optical and MIR classifications are broadly
consistent with each other (i.e., type 1 AGN usually show strong continuum and silicate emission, 
type 2 show also strong continuum and silicate absorption, and starbursts show
strong PAH features on top of a fainter continuum), but there are exceptions. 
In particular, more than 1/4 of Seyfert 1s and QSOs show significant silicate absorption,
and many sources classified as AGN in the optical are starburst dominated in the MIR.
Apperture effects may be at least in part responsible for this, since the IRS spectrum of a galaxy nucleus include
a far larger share of the emission from the galaxy compared to the corresponding optical spectrum. 

In the lower luminosity Seyferts, the host galaxy can contribute by a large fraction
to the luminosity within the IRS slit, and composition of the (AGN-dominated) nuclear spectrum and 
(starburst-dominated) host galaxy spectrum can produce a silicate feature in absorption.
This interpretation is 
supported by the strong PAH features observed in some Seyferts, that make them qualify as 
starburst-dominated in the MIR. 
Nevertheless, significant silicate absorption is also present in
the spectrum of several QSOs, where the contribution from the host galaxy to the MIR luminosity 
is expected to be small. Thus at least in some cases the silicate absorption feature may be caused by
obscuration of the AGN emission, be it in a dusty torus or in the host galaxy.

On the other hand, there are also 15 Seyfert 2s and QSO2s (1/6 of the total) with the silicate 
in emission, as meassured by \tausil (12 of them also confirmed by \ssil $>$ 0). 

These apparent discrepancies, however, can not be regarded as evidence against the unified model of AGN; in fact,
such behaviour is predicted by both smooth and clumpy circumnuclear dust distribution models. 
Smooth models predict silicate absorption in compact, high opacity face-on tori \citep{Fritz06}, and also allow for silicate emission in type 2 AGN
\citep[e.g.][]{Hatziminaoglou08}. Clumpy models require a high number of clouds along the equator and 
high optical depth to produce silicate absorption in type 1
AGN \citep{Nenkova08}, while some parameter combinations could give rise
to silicate emission in type 2 AGN \citep[e.g.][]{Nikutta09}.

We therefore conclude that a silicate emission or absorption feature is not a 
reliable indicator, on its own, of the optical spectral type.

There are also four LINERs with silicate emission, three of them show broad Balmer lines, and a strong
silicate peak, while the remaining one is a radiogalaxy with a weaker silicate feature.

Almost all optical starbursts are also classified as starbursts in the MIR, but the opposite is not true:
roughly half of MIR starbursts belong to some of the optical AGN classes. These sources
are mostly nearby galaxies harbouring low luminosity AGN, that can be outshined by the galaxy in the 
large IRS aperture. When taking into account only objects at $z$ $>$ 0.1, from 98 MIR\_SB sources 54 are
classified as starbursts in the optical, while 31 have LINER cores and only 13 are Seyferts or QSOs.

\section{Average templates}

A number of synthetic galaxy templates have been obtained from the average spectrum of populations 
matching certain criteria, shown in Table \ref{tabla_templates}. 

We obtain average templates for Seyfert 1, Seyfert 1.X, Seyfert 2, LINER, QSO, QSO2, starburst, ULIRGs and SMGs.
ULIRGs with LINER or starburst cores are discarded when composing the LINER and starburst templates, that
include only sources with $\nu$L$_\nu$(7\uu) $<$ 10$^{44}$ erg s$^{-1}$.
Three additional
templates are generated for sources classified in the MIR as dominated by starburst (MIR\_SB), 
unobscured AGN (MIR\_AGN1) and obscured AGN (MIR\_AGN2).

\begin{figure*}
\begin{center}
\includegraphics[width=17cm]{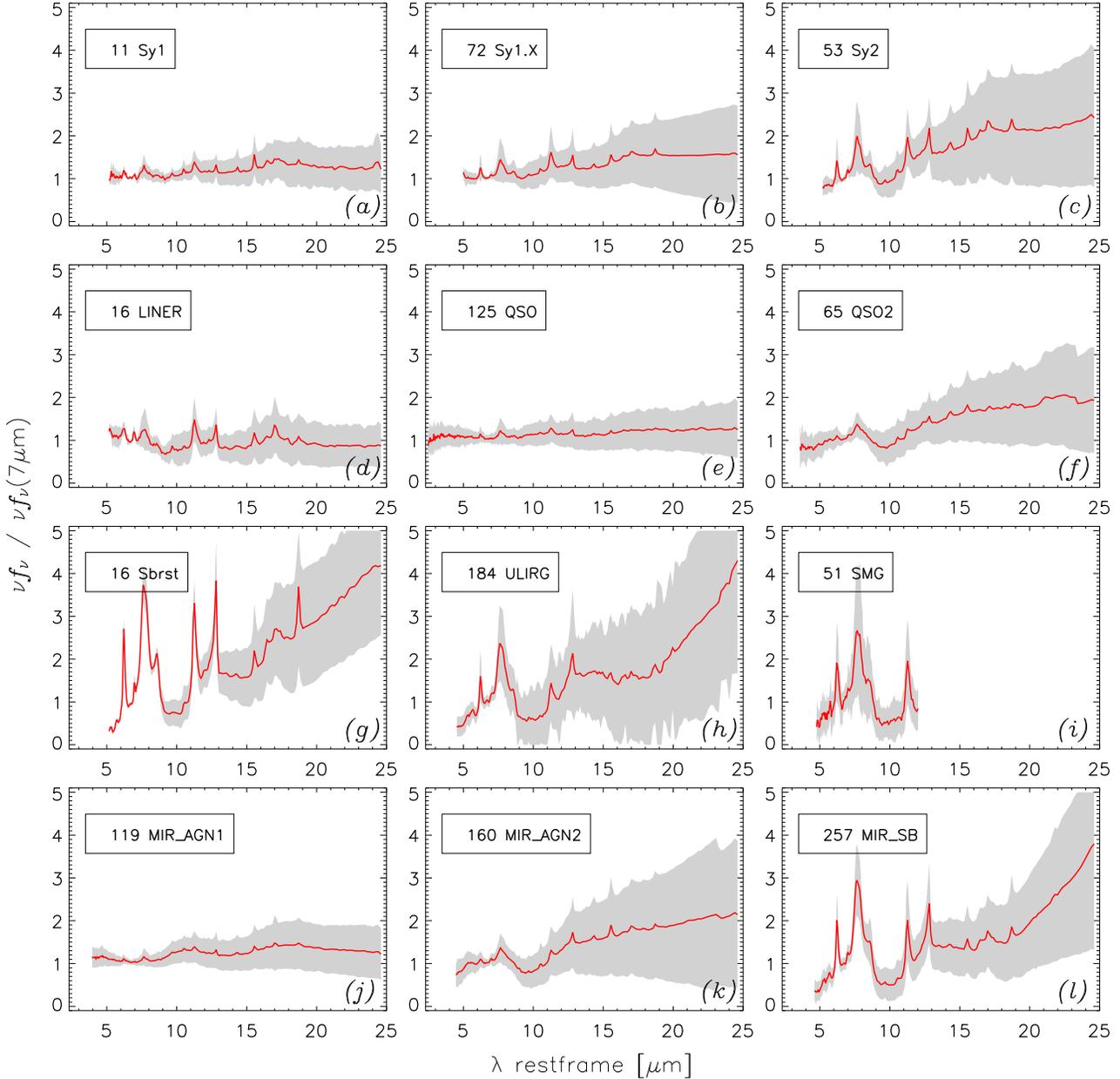}
\end{center}
\caption[Templates]{Composite templates for subsamples of the ATLAS sample. In the first three rown, sources are selected based on their
optical type and MIR luminosity, except for the ULIRG and SMG templates (see text). The bottom row shows average templates of
the sources grouped according to the MIR classification. \label{templates3x4}}
\end{figure*}

The individual spectra are normalized at 7 \um resframe and resampled to a common grid of wavelength values.
The highest and lowest flux values in each wavelength bin are discarded, and the remaining ones are averaged
with the arithmetic mean to obtain the template flux.
The 12 templates obtained are shown in Figure \ref{templates3x4}.
Shaded areas indicate the 1-$\sigma$ dispersion of individual
spectra, which is higher at longer wavelengths because of slope variations among sources of the same type.
These templates are available as ASCII tables with columns for rest frame wavelength, normalized
average flux density, 1-$\sigma$ dispersion in flux density and number of sources contributing to the
average for each wavelength bin.

The QSO template $(e)$ is flat and almost featureless, while the type 2 QSOs $(f)$ have on average a redder continuum
and a somewhat pronounced silicate feature in absorption. 
The equivalent with of the PAH features in the QSO and QSO2 templates are much lower compared to
those present in the average Sy1 $(a)$ and Sy2 $(c)$ templates, in spite of the high star formation rates derived from 
both the PAH features \citep[e.g.][]{Hernan-Caballero09} and FIR luminosities \citep[e.g.][]{Hatziminaoglou10}, because the strong 
continuum emission from the AGN masks the underlying emission from the host galaxy. 
The LINER template $(d)$ includes only nearby sources and its low PAHs equivalent widths are indicative of weak
circumnuclear star formation.
Its 5--9 \um continuum is bluer than either the starburst or Seyfert templates because contribution from starlight
is not negligible compared to the AGN emission. 
Complementary templates are generated for all sources classified as ULIRG $(h)$ in the literature (irrespective of their
optical spectral type) as well as the SMGs $(i)$. The latter has a very limitted spectral coverage, as all the objects
composing the SMG sample are high-redshift objects (with an average redshift $\langle z \rangle$ = 1.93).

The MIR AGN average templates $(j, k)$ also show interesting characteristics: the objects selected to have silicate
in emission at 10 \um have a flat spectrum with the silicate feature at 18 \um clearly in emission, as well.
The objects selected to have the silicate feature in absorption have a redder continuum and show no feature
at 18 \uu, be it in emission or in absorption. In a detailed comparative study between smooth and clumpy
dust distributions, Fetre et al (in prep) show that such silicate features' behaviour is compatible with both
clumpy and smooth dust distributions and hence we can not derive any constraints from the average templates.

Interestingly, the average starburst template constructed from objects selected as such in the optical $(g)$
shows more prominent PAH features and a redder continuum than the equivalent MIR template $(l)$. This difference is most likely due to the
different nature of the objects composing the two samples. The optical sample consists of low redshift
starburst galaxies alone, while the MIR template has a higher average redshift, includes many ULIRGs as well as
objects with significant AGN contribution.

\section{Conclusions} 

We have collected a sample of 739 active and starburst galaxies with {\it Spitzer}/IRS spectra from the literature.
The sample, though not statistical,
comprises a variety of objects, namely type 1, 2 QSOs and Seyfert galaxies, intermediate Seyfert galaxies,
ULIRGs, starburst and sub-mm galaxies, with a large
span of spectral properties such as PAH and silicate features, continua
and distinct physical characteristics, and is the largest of its kind compiled to date. MIR properties were
computed in a concise way for all objects and can thus be directly compared among the various types of objects.

We find that the silicate strength, \ssilu, a meassurement of the intensity of the 10 \um silicate feature that takes 
positive (negative) values when the feature is in emission (absorption), is well aproximated 
by a ``three band'' color index, \csilu, that is usually easier to obtain, except for sources with strong PAH emission
that contaminates the flux at 7 \uu, causing \csil to consistently lie short of \ssilu.
A more elaborate meassurement of dust obscuration, \tausilu, obtained from the scaling factor 
applied to the extinction law in a spectral decomposition analysis, shows also a tight correlation with \ssilu, 
but with different slopes for sources with saturated (\tausil $>$ 3) and unsaturated (\tausil $<$ 3) silicate features.

We find that the ratio of the PDR component to the total luminosity of each object, integrated between 5 and 15 \um restframe, 
$r_{\rm{PDR}}$, and \tausil, both obtained from the spectral decomposition, provide better 
AGN versus SB and obscuration diagnostics than the usual PAH EW and \ssil in sources with sufficient spectral coverage, 
as they do not depend on a single spectral feature and thus are less vulnerable to noise.
Most QSOs and Seyfert1 galaxies show \tausil values clustered around zero and $r_{\rm{PDR}}$ values below 0.15,
while starburst galaxies have large $r_{\rm{PDR}}$ values and moderate to deep silicate absorption
(higher in the starburst ULIRGs than in lower luminosity starbursts).
Type 2 QSOs and Seyfert 2s exhibit low $r_{\rm{PDR}}$ values and usually silicate absorption,
sometimes strong, while
LINERs can have almost any combination of parameters, due to their nature and definition.

We then derive a simple classification of the sources, based on the values of $r_{\rm{PDR}}$ and \tausilu.
Objects with $r_{\rm{PDR}} < 0.15$ are considered to be AGN-dominated, while $r_{\rm{PDR}} > 0.15$ indicates
starburst-dominated sources, and the AGN-dominated subsample is further divided into silicate emitters and
absorbers (\tausil $>$ 0 and \tausil $<$ 0, respectively).
Comparison of this simple MIR classifications with the optical classifications in the literature
is not straight forward, since the much wider apperture of the IRS slit with respect to optical 
spectroscopy allows for a larger fraction of the host galaxy in the MIR spectrum of resolved sources.
Nevertheless, optical and MIR classifications are broadly consistent.
Apperture effects may explain the larger fraction of starburst-dominated sources in the MIR relative to optical
classifications, and the composition of nuclear and host galaxy spectra might account for some of the type 1
Seyfert galaxies that exhibit silicate absorption, even if in higher luminosity QSOs a source of obscuration (whether in
a circumnuclear dusty torus or in the host galaxy) is required to explain the cases of silicate absorption.
Since up to 1/4 of the type 1 AGN show silicate absorption and 1/6 of type 2 AGN show silicate 
emission, we conclude the silicate feature is not, on its own, a reliable indicator of the AGN optical type.

All spectra in the sample, the average templates, as well as ancillary data, including (but not limited to) optical, IR and
radio photometry, redshifts, and MIR spectral meassurements are made available to the public at the ATLAS  website: http://www.denebola.org/atlas.

\section*{Acknowledgements}

This work is based on observations made with the \textit{Spitzer Space
Telescope}, which is operated by the Jet Propulsion Laboratory, Caltech
under NASA contract 1407.\\
This work made use of the software TOPCAT (http://www.star.bris.ac.uk/$\sim$mbt/topcat/)
developed by M. Taylor.\\
We wish to thank Jorge A. P\'erez Prieto for his kind assistance with the ATLAS web site,
and the anonymous referee for the very useful comments and suggestions.

\bsp
\onecolumn

\begin{deluxetable}{llll} 
\tabletypesize{\scriptsize}
\tablecaption{IRS observations log\label{IRS_observations_data}}
\tablewidth{0pt}
\tablehead{
\colhead{Sample} & \colhead{IRS modules} & \colhead{Spitzer Archive PID(s)} & \colhead{Comments}}
\startdata
Brand et al. 2008, ApJ, 680, 119                        & SL1,LL2,LL1    &  15                        &   16 optically faint QSOs with X-ray detection  \\
Brandl et al. 2006 , ApJ, 653, 1129                     & SL,LL          &  14                        &   22 nearby starbursts                          \\
Buchanan et al. 2006, AJ, 132, 401                      & SL2-LL1 (map)  &  3269                      &   87 nearby 12um-selected Seyfert galaxies      \\
Cao et al. 2008, MNRAS, 390, 336                        & SL,LL          &  various                   &   19 ultraluminous IR QSOs                      \\
Dasyra et al. 2009, ApJ, 701, 1123                      & SL1,LL2,LL1*   &  20629                     &   150 24\um selected IR-luminous galaxies        \\
Deo et al. 2007, ApJ, 671, 124                          & SL,LL          &  3374                      &   12 Seyfert 1.8 and 1.9 galaxies                 \\
Farrah et al. 2009, ApJ, 696, 2044                      & SL1,LL2,LL1*   &  30364                     &   16 optically faint 70 \um sources               \\
Haas et al. 2005, A\&A, 442, L39                        & SL,LL          &  3349                      &   7 quasars and 7 radio galaxies                  \\
Hao et al. 2005, ApJ, 625, L75                          & SL,LL          &  14                        &   5 PG quasars                                  \\
Hern\'an-Caballero et al. 2009, MNRAS, 395,  1695       & SL,LL          &  3640                      &   69 15um-selected IR-luminous galaxies         \\
Hiner et al. 2009 , ApJ, 706, 508                       & SL,LL          &  20083                     &   6 type 1 and 7 type 2 MIR selected quasars \\
Imanishi et al. 2007, ApJ, 171, 72                      & SL,LL          &  105,2306,3187             &   buried AGN in IR galaxies                     \\
Imanishi et al. 2009, ApJ, 694, 751                     & SL,LL          &  105,30407,20589           &   20 buried AGN in IR galaxies                  \\
Imanishi et al. 2010, ApJ, 709, 801                     & SL,LL          &  50008,3187,30407          &   17 nearby ULIRGs                              \\
Lacy et al. 2007, ApJ, 669, 61                          & SL1,LL2        &  20083                     &   6 type 2 quasars                              \\
Leipski et al. 2009, ApJ, 701, 891                      & SL,LL          &  20525                     &   25 FR-I radio galaxies                        \\
Lutz et al. 2008, ApJ, 684, 853                         & LL             &  30314                     &   12 z$\sim$2 mm-bright type 1 quasars               \\
Maiolino et al. 2007, A\&A, 468, 979                    & LL1            &  20493                     &   4 high luminosity quasars 2$<$z$<$3.5             \\
Men\'endez-Delmestre et al. 2009, ApJ, 699, 667         & LL1,LL2*       &  20081                     &   24 submm galaxies                             \\
Murphy et al. 2009, ApJ, 698, 1380                      & SL1,LL2,LL1    &  20456                     &   22 galaxies in GOODS-N                        \\
Netzer et al. 2007 ApJ, 666, 806                        & SL,SH,LH       &  3187                      &   23 Palomar Green QSOs                         \\
Polletta et al. 2008, ApJ, 675, 960                     & SL1,LL2,LL1    &  various                    &   21 obscured AGNs                              \\
Pope et al. 2008, ApJ, 675, 1171                        & LL2,LL1*       &  20456,20081                &   13 submm galaxies                             \\
Shi et al. 2006, ApJ, 653, 127                          & SL,LL          &  various                    &   68 AGN                                        \\
Siebenmorgen et al. 2005, A\&A, 436, L5                 & SL,LL          &  20231                      &   3C 249.1, 3C 351                              \\
Sturm et al. 2005, ApJ, 629, L21                        & SL,SH,LH       &  3237                      &   NGC 3998                                      \\
Sturm et al. 2006, ApJ, 642, 81                         & SL,LL*         &  3223                      &   7 X-ray selected type 2 quasars               \\
Tommasin et al. 2008, ApJ, 676, 836                     & SH,LH          &  30291                     &   29 Seyfert galaxies                           \\
Valiante et al. 2007, ApJ, 660, 1060                    & SL,LL*         &  3241                      &   9 submm galaxies                              \\
Weedman et al. 2005, ApJ, 633, 706                      & SL,LL,SH,LH    &  14                        &   8 AGN                                         \\
Weedman et al. 2006a, ApJ, 651, 101                     & SL1,LL2,LL1*   &  12,15                     &   24 optically faint 24 \um selected AGN         \\
Weedman et al. 2006b, ApJ, 653, 101                     & SL1,LL2,LL1*   &  15                         &   11 AGN and 9 starbursts                       \\
Weedman et al. 2009, ApJ, 693, 370                      & SL,LL*         &  40038,20128,20083,40539   &   24 AGNs and some starbursts                   \\
Willett et al. 2010, ApJ, 713, 1393                     & SL,LL,SH,LH    &  30515,50591,105           &   8 compact radio galaxies                      \\
Wu et al. 2009, ApJ, 701, 658                           & SL,LL (map)    &  various                    &   103 Seyfert galaxies                          \\
Yan et al. 2007, ApJ, 658, 778                          & SL1,LL2,LL1*   &  3748                      &   52 24 \um seleted sources                      \\
Zakamska et al. 2008, AJ, 136, 160                      & SL1,LL2        &  105,3163                  &   12 QSO2s from SDSS                            \\
\enddata
\tablenotetext{*}{not all targets observed in all modules}
\end{deluxetable}

\begin{deluxetable}{l c c r} 
\tabletypesize{\scriptsize}
\tablecaption{monochromatic luminosity meassurement parameters\label{tabla_nuLnu}}
\tablewidth{0pt}
\tablehead{
\colhead{band name} & \colhead{$\lambda_{rest}$} &\colhead{bandwidth} &\colhead{N$_{sources}$} \\
\colhead{}          & \colhead{[\um]}            &\colhead{[\um]}     &\colhead{}         }
\startdata
K            &    2.2     &    0.2    &   29  \\
3.6          &    3.6     &    0.2    &  102  \\
5.6          &    5.55    &    0.5    &  485  \\
7            &    7.0     &    0.5    &  630  \\
9            &    9.0     &    0.5    &  656  \\
10           &    10.0    &    1.0    &  624  \\
12           &    12.0    &    1.0    &  565  \\
14           &    14.0    &    1.5    &  508  \\
15           &    15.0    &    1.5    &  489  \\
18           &    18.0    &    1.5    &  430  \\
20           &    20.0    &    1.5    &  397  \\
25           &    25.0    &    2.0    &  352  \\
\enddata
\end{deluxetable}

\begin{deluxetable}{l c c} 
\tabletypesize{\scriptsize}
\tablecaption{PAH meassurement parameters\label{tabla_pahs}}
\tablewidth{0pt}
\tablehead{
\colhead{} & \colhead{6.2\um PAH} &\colhead{11.3\um PAH}}
\startdata
central wavelength &  6.25 & 11.3 \\
integration band   &  6.0-6.5 &  11.05-11.55 \\
continuum intervals & 5.8-6.0,6.5-6.7 & 10.75-11.0,11.65-11.9\\
assumed FWHM & 0.2 & 0.2 \\    
\enddata
\end{deluxetable}

\begin{deluxetable}{c  c c  c c  c c} 
\tabletypesize{\scriptsize}
\tablecaption{Bias in decomposition parameters caused by incomplete spectral coverage\label{test_decomp}}
\tablewidth{0pt}
\tablehead{
\colhead{} & \multicolumn{2}{c}{5.0--13.5\uu} & \multicolumn{2}{c}{5.0--12.0\uu} & \multicolumn{2}{c}{5.0--10.5\uu} \\
\colhead{} & \colhead{mean} & \colhead{$\sigma$} & \colhead{mean} & \colhead{$\sigma$} & \colhead{mean} & \colhead{$\sigma$}}
\startdata
$\Delta r_{\rm{AGN}}$ &  0.0029  & 0.017 &  0.0426 & 0.051 &  0.0590 & 0.118 \\
$\Delta r_{\rm{HII}}$ & -0.0081  & 0.034 & -0.0699 & 0.098 & -0.0499 & 0.193 \\
$\Delta r_{\rm{PDR}}$ &  0.0052  & 0.029 &  0.0272 & 0.065 & -0.0092 & 0.096 \\
$\Delta$\tausil       & -0.0005  & 0.141 &  0.0027 & 0.560 &  0.2723 & 0.820 \\
\enddata
\end{deluxetable}

\begin{deluxetable}{l r@{}l r@{}l r@{}l r@{}l r@{}l r@{}l } 
\tabletypesize{\scriptsize}
\tablecaption{Comparison of optical and MIR classifications\label{optical_MIR_table}}
\tablewidth{0pt}
\tablehead{
\colhead{} & \multicolumn{2}{c}{MIR\_AGN1} & \multicolumn{2}{c}{MIR\_AGNx} &
\multicolumn{2}{c}{MIR\_AGN2} & \multicolumn{2}{c}{MIR\_AGN} &
\multicolumn{2}{c}{MIR\_COMP} & \multicolumn{2}{c}{MIR\_SB} }
\startdata
Sy1 &  7 & &  0 & &  3 & &  0 & &  0 & &  1 & (2) \\
Sy1.X & 26 & &  6 & & 18 & &  1 & &  0 & & 20 & (24) \\
Sy2 &  7 & &  1 & & 25 & (26) &  0 & &  1 & & 19 & (24) \\
LINER &  4 & &  0 & &  4 & &  0 & &  0 & &  9 & \\
QSO & 61 & (63) &  8 & & 23 & (26) & 19 & &  2 & &  5 & (8) \\
QSO2 &  8 & &  2 & & 31 & (34) &  7 & &  4 & &  5 & (7) \\
Sbrst/HII &  0 & (2) &  0 & &  5 & (12) &  1 & &  7 & & 74 & (89) \\

\enddata
\end{deluxetable}

\begin{deluxetable}{l c c c c c c l} 
\tabletypesize{\scriptsize}
\tablecaption{Composite spectral templates\label{tabla_templates}}
\tablewidth{0pt}
\tablehead{
\colhead{name} & \colhead{Nsources} & \colhead{$z_{min}$} & \colhead{$\langle z \rangle$} & \colhead{$z_{max}$} & 
\colhead{$\lambda_{min}$ [\uu]} & \colhead{$\lambda_{max}$ [\uu]} & \colhead{comments}   }
\startdata
  Sy1 &  11 &   0.002 &   0.041 &   0.205 &     5.2 &    24.6 &                      Seyfert 1 with $\nu$L$\nu$(7\uu) $<$ 10$^{44}$ erg s$^{-1}$\\
  Sy1x &  72 &   0.003 &   0.091 &   0.371 &     5.0 &    24.6 &                                  intermediate Seyfert types (1.2, 1.5, 1.8, 1.9)\\
  Sy2 &  53 &   0.003 &   0.045 &   1.140 &     5.2 &    24.6 &                      Seyfert 2 with $\nu$L$\nu$(7\uu) $<$ 10$^{44}$ erg s$^{-1}$\\
LINER &  16 &   0.001 &   0.034 &   0.322 &     5.2 &    24.6 &                          LINER with $\nu$L$\nu$(7\uu) $<$ 10$^{44}$ erg s$^{-1}$\\
  QSO & 125 &   0.020 &   1.092 &   3.355 &     2.5 &    24.6 &             QSO1 and Seyfert 1 with $\nu$L$\nu$(7\uu) $>$ 10$^{44}$ erg s$^{-1}$\\
 QSO2 &  65 &   0.031 &   1.062 &   3.700 &     3.6 &    24.6 &             QSO2 and Seyfert 2 with $\nu$L$\nu$(7\uu) $>$ 10$^{44}$ erg s$^{-1}$\\
Sbrst &  16 &   0.001 &   0.091 &   1.316 &     5.2 &    24.6 &               Starburst or HII with $\nu$L$\nu$(7\uu) $<$ 10$^{44}$ erg s$^{-1}$\\
     ULIRG & 184 &   0.018 &   0.730 &   2.704 &     4.5 &    24.6 &                                            ULIRG (low and high redshift sources)\\
       SMG &  51 &   0.557 &   1.869 &   3.350 &     4.8 &    12.0 &                                                           Submillimiter Galaxies\\
  MIR\_AGN1 & 119 &   0.002 &   0.455 &   2.190 &     4.0 &    24.6 &                                          MIR selected AGN with silicate emission\\
  MIR\_AGN2 & 160 &   0.002 &   0.549 &   2.470 &     4.5 &    24.6 &                                        MIR selected AGN with silicate absorption\\
    MIR\_SB & 257 &   0.001 &   0.413 &   2.000 &     4.6 &    24.6 &                                                          MIR selected starbursts\\

\enddata
\end{deluxetable}

\end{document}